\documentclass[letterpaper, 11 pt]{article}                                                          

\usepackage{graphicx}                                                          
\usepackage{amsmath,amssymb}
\usepackage[noadjust]{cite}                                                     

\date{}

\title{\Huge
Is Lipschitz Continuity Preserved under Sampled-Data Discretization?}

\author{Masoud Abbaszadeh
\thanks{M. Abbaszadeh is with GE Global Research, NY, USA, {\tt\small e-mail: masoud@ualberta.net}}}

\begin{document}

\maketitle \thispagestyle{empty} \pagestyle{empty}

\begin{abstract}
Usually, given a continuous-time nonlinear model, a closed form
solution for an exact discretization cannot be found explicitly,
originating the need of approximating discrete-time models.
This note studies the preservation of the Lipschitz continuity under approximate discretizations.
\end{abstract}


\section{Introduction}
The theory of nonlinear sampled data systems has still a long way
ahead to be well developed. An interesting and important path for
further research is to extend previous results using Euler discretization to higher
order approximations. In control systems, the
discretization is often under the Zero-Order Hold assumption, i.e. the control input is assumed to be constant during
the sampling intervals $\left[kT, (k+1)T\right)$, where $T$ is the sampling time. Under this
assumption,  the system becomes autonomous within the sampling interval and
  thus it is amenable to a rigorous mathematical treatment rooted in
  the well-known Taylor-Lie series theory for nonlinear autonomous
  ODEs, widely used in this context. In particular, the analytical solution of the systems is
  expandable in a uniformly convergent Taylor series within the
  sampling interval and the resulting coefficients can be easily
  obtained by taking successive partial derivatives of the
  right-hand side of systems model.
  On the other hand, there is a large body of literature for control and estimation of nonlinear systems satisfying a Lipschitz continuity condition.
  See for example \cite{Rajamani5,abbaszadeh2008robust, abbaszadeh2007robust, Raghavan, abbaszadeh2006robust,Xu1, Xu2, Xu3, Xu4,Hammouri,Lu,Gao,
abbaszadeh2010nonlinear, abbaszadeh2008lmi, Thau1, abbaszadeh2010dynamical, abbaszadeh2010robust2,Rajamani2, deSouza1, deSouza2, abbaszadeh_phdthesis,
abbaszadeh2012generalized,Abbaszadeh5} and the references therein, for details of the approach and applications to
control and filtering of different classes of nonlinear systems. The significance of this condition is that it guarantees the existence and uniqueness of the solution of the nonlinear systems. Also, it provides a mathematically tractable framework to apply Lyapunov stability theory and establish stability and performance conditions in the form of Riccati equations or LMIs.

This note studies whether Lipschitz continuity is preserved under approximate discretizations of nonlinear system using a zero order hold, for both standard two-sided Lipschitz condition and its extended one-sided version. Analytical expressions are given relating the Lipschitz constants of discretized systems to their continuous-time values.

\section{Problem Statement}
We consider the following continuous-time system
\begin{align}
\dot{x}&=Ax+f(x,u) \label{con1}\\
y&=Cx \label{con2}
\end{align}
where $x\in {\mathbb R} ^{n} ,u\in {\mathbb R} ^{m} ,y\in {\mathbb
R} ^{p} $. The control input is assumed to be constant during
the sampling intervals $\left[kT, (k+1)T\right)$ (zero-order hold
assumption). The family of exact discretization is:
\begin{align}
x_{k+1}&=A_{d}^{e} x_{k}+F_{T}^{e} (x_{k},u_{k}) \\
y_{k}&=C_{d} x_{k}.\notag
\end{align}
Index {\it T} means the discretization is dependent to the sampling
time. Often, the exact discretization is not available. However, it is realistic to assume that a family of
approximate discrete-time models is available
\begin{align}
x^{a}_{k+1}&=A_{d}^{a}x^{a}_{k}+F_{T}^{a}(x^{a}_{k},u_{k}) \\
y_{k}&=C_{d}x^{a}_{k}.\notag
\end{align}

Before stating the problem, we need to refer to the following two definitions. \\

\textbf{Definition 1. \cite{abbaszadeh2010nonlinear}} The system (\ref{con1})-\eqref{con2}
is said to be \emph{locally
Lipschitz} in a region $\mathcal{D}$ including the origin with
respect to $x$, uniformly in $u$, if there exist a constant $\gamma_c>0$ satisfying:
\begin{eqnarray}
\|f(x_{1},u^{*})-f(x_{2},u^{*})\|\leqslant \gamma_{c}\|x_{1}-x_{2}\|
\hspace{3mm}\forall \, x_{1},x_{2}\in \mathcal{D},\label{Lip}
\end{eqnarray}
where  $u^{*}$ is any admissible control signal. The smallest constant $\gamma_{c}>0$ satisfying (\ref{Lip}) is known as the
\emph{Lipschitz constant}.
The region $\mathcal{D}$ is the \emph{operational region} or our \emph{region of interest}. If the condition
(\ref{Lip}) is valid everywhere in $\mathbb{R}^{n}$, then the function is said to be globally Lipschitz.\\
\newline
An extension of this class are the so-called \emph{one-sided} Lipschitz systems.
The following definition introduces one-sided Lipschitz functions.\\
\newline
\textbf{Definition 2. \cite{abbaszadeh2010nonlinear}} The system (\ref{con1})-\eqref{con2} is said to be
\emph{one-sided Lipschitz} if there exist $\rho_c \in \mathbb{R}$ such that
$\forall \, x_{1},x_{2}\in \mathcal{D}$
\begin{eqnarray}
\left<f(x_{1},u^{*})-f(x_{2},u^{*}), x_{1}-x_{2}\right> \
\leqslant \rho_{c} \|x_{1}-x_{2}\|^{2},\label{con3}
\end{eqnarray}
where $\rho_{c} \in \mathbb{R}$ is called the \emph{one-sided Lipschitz constant}. As in the case of Lipschitz functions,
the smallest $\rho_c$ satisfying \eqref{con3} is considered as the one-sided Lipschitz constant. Similar to the Lipschitz property, the one-sided Lipschitz property might be local or global.\\

Note that while the Lipschitz constant must be positive,
the one-sided Lipschitz constant can be positive, zero or even negative \cite{abbaszadeh2010nonlinear}.
For any function $f(x,u)$, we have:
\begin{align}
&|\left<f(x_{1},u^{*})-f(x_{2},u^{*}), x_{1}-x_{2}\right>| \notag\\
&\hspace{20mm} \leqslant \|f(x_{1},u^{*})-f(x_{2},u^{*})\|\|x_{1}-x_{2}\|\notag\\
&\text{and if $f(x,u)$ is Lipschitz, then:} \ \ \ \ \ \leqslant  \gamma_{c} \|x_{1}-x_{2}\|^{2}.\notag
\end{align}
Therefore, any Lipschitz function is also one-sided Lipschitz. The converse, however, is not true.
For Lipschitz functions, we have
\begin{align}
-\gamma_{c} \|x_{1}-x_{2}\|^{2} &\leqslant \left<f(x_{1},u^{*})-f(x_{2},u^{*}), x_{1}-x_{2}\right> \leqslant \gamma_{c} \|x_{1}-x_{2}\|^{2},\notag
\end{align}
which is a \emph{two-sided} inequality v.s. the \emph{one-sided} inequality in \eqref{con3}.
The Lipschitz constants can be bounded by the norms of the Jacobian \cite{Marquez} or computed through numerical optimization \cite{Wood}. See \cite{abbaszadeh2010nonlinear} for further details. \\

Assuming the continuous-time system is Lipschitz or one-sided Lipschitz, the purpose of this note is to study the conditions under which these properties are preserved under zero order hold discretization of the nonlinear system.

\section{Lipschitz Conditions under ZOH Discretization}
Under the ZOH assumption, similar to the approach given in \cite{Kazantzis1999763}, we have:
\begin{equation}
\begin{split}
x(k+1)&=x(k)+{\sum_{l=1}^{\infty}\frac{T^{l}}{l!}\frac{d^{l}x}{dt^{l}}}|_{t_{k}}\\
&=x(k)+\sum_{l=1}^{\infty}\frac{T^{l}}{l!}\frac{d^{l-1}}{dt^{l-1}}[Ax+f(x,u)]|_{t_{k}}\\
&=x(k)+\sum_{l=1}^{\infty}\frac{T^{l}}{l!}[A\frac{d^{l-1}x}{dt^{l-1}}+\frac{d^{l-1}}{dt^{l-1}}f(x,u)]|_{t_{k}}.
\end{split}
\end{equation}
where {\large\begin{equation} \left\{
  \begin{array}{l}
    \frac{d}{dt}f(x,u)=\frac{\partial f}{\partial x}\cdot
\frac{dx}{dt}+\frac{\partial f}{\partial u}\cdot \frac{du}{dt}\\
    \frac{d^{n}}{dt^{n}}f(x,u)=
\frac{d}{dt}[\frac{d^{n-1}}{dt^{n-1}}f(x,u)], \ \ \ n \geq 2 \\
  \end{array}
\right.
\end{equation}}
Under the ZOH assumption, $\frac{du}{dt}=0$ in each sampling
interval and thus:
\begin{equation}
\begin{split}
x(k+1)&=x(k)+\sum_{l=1}^{\infty}\frac{T^{l}}{l!}[A\frac{d^{l-1}x}{dt^{l-1}}+\frac{d^{l-1}}{d
t^{l-1}}f(x,u)]|_{t_{k}} \\
\frac{d}{dt}f(x,u)&=\frac{\partial f}{\partial x}\cdot
\frac{dx}{dt}, \ \ \ \ \ \ \ \frac{d^{n}f}{dt^{n}}=
\frac{d}{dt}(\frac{d^{n-1}f}{dt^{n-1}}), \ n \geq 2. \label{taylor2}
\end{split}
\end{equation}
The first order approximation, ($l = 1$) leads to the well-known Euler
approximate model.
\subsection{First Order Discrete Approximation (the Euler Method)}
We first analyse Lipschitz conditions under the Euler discritization. This is a trivial case,
in which Lipschitz continuity of the discretized system is established following the properties of the inner-product spaces.
Yet this is a very important case, since it has significant practical applications due to
its computational simplicity. For the Euler approximation we have
\begin{align}
x(k+1)&=x(k)+T[Ax(k)+f(x_{k},u_{k})], \\
A_{d}^{a}&=I+AT,\ C_{d}=C,  \notag\\
F_{T}^{a}(x^{a}_{k},u_{k})&=T f(x^{a}_{k},u_{k}).\notag
\end{align}
\subsubsection{Lipschitz Continuity}
As seen, under the Euler discretization, the structure of the nonlinear function
is preserved and is just scaled by the sampling time. Therefore,
\begin{align}
\|F_{T}^{a}(x_{1},u^{*})-F_{T}^{a}(x_{2},u^{*})\| = T\|f(x_{1},u^{*})-f(x_{2},u^{*})\|\leqslant T\gamma_{c}\|x_{1}-x_{2}\|
\hspace{3mm}\forall \, x_{1},x_{2}\in \mathcal{D},\label{Lip2}
\end{align}
where, $x_{1}$ and $x_{2}$ are any two points in the operating space.
It is clear that the Lipschitz continuity is preserved under Euler approximation. The Lipschitz constant of the Euler
approximate model is $\gamma _{d} =T\gamma _{c}$.\\
\subsubsection{One-Sided Lipschitz Continuity}
A similar argument can be made for the one-sided Lipschitz property.
\begin{align}
\left<F_{T}^{a}(x_{1},u^{*})-F_{T}^{a}(x_{2},u^{*}), x_{1}-x_{2}\right> &= T\left<f(x_{1},u^{*})-f(x_{2},u^{*}), x_{1}-x_{2}\right> \notag\\
&\leqslant T\rho_{c} \|x_{1}-x_{2}\|^{2}, \hspace{3mm}\forall \, x_{1},x_{2}\in \mathcal{D}.\label{Lip3}
\end{align}
This means that the Euler approximate model in one-sided Lipschitz with one-sided Lipschitz contact $\rho_{d} =T\rho_{c}$.

So, in summary, using the Euler discretization, both two-sided and one-sided Lipschitz constant are just linearly scaled by the sampling time.
\subsection{Second Order Discrete Approximation}
The Taylor expansion of the second order discrete model gives:
\begin{align}
  x(k+1) &= \underbrace{\left(I+AT+\frac{T^2}{2}A^2\right)}_{A_{d}^{a}}x(k)+\underbrace{\left(T+\frac{T^2}{2}A\right)f(x,u)+\frac{T^2}{2}\left(\frac{\partial f}{\partial x}Ax+\frac{\partial f}{\partial x}f(x,u)\right)}_{F_{T}^{a}(x,u)}
\end{align}
\subsubsection{Lipschitz Continuity}
Based on the above:
\begin{align}
&\|F_{T}^{a}(x_{1},u^{*})-F_{T}^{a}(x_{2},u^{*})\|= \\\notag
&\Bigg\|\left(T+\frac{T^2}{2}A\right)f(x_1,u^{*})+\frac{T^2}{2}\left({\frac{\partial f}{\partial x}}_{|x = x_1}Ax_1+\frac{\partial f}{\partial x}_{|x = x_1}f(x_1,u^{*})\right)\\ \notag
&- \left(T+\frac{T^2}{2}A\right)f(x_2,u^{*})-\frac{T^2}{2}\left({\frac{\partial f}{\partial x}}_{|x = x_2}Ax_2+\frac{\partial f}{\partial x}_{|x = x_2}f(x_2,u^{*})\right)\Bigg\|\\ \notag
&< T \|f(x_1,u^*)-f(x_2,u^*)\|+\frac{T^2}{2}\overline{\sigma}(A)\alpha\|x_1-x_2\|+\frac{T^2}{2}(\overline{\sigma}(A)+\alpha)\|f(x_1,u^*)-f(x_2,u^*)\|,
\end{align}
where $\overline{\sigma}(A)$ is the maximum singular value or the induced 2-norm of $A$ and $\alpha$ is the supremum of the norm of the Jacobian of $f(x,u)$ over the operating region. Using the Lipschitz continuity condition we get:
\begin{align}
 &\|F_{T}^{a}(x_{1},u^{*})-F_{T}^{a}(x_{2},u^{*})\| < \left(T\gamma_c+\frac{T^2}{2}\overline{\sigma}(A)(\alpha+\gamma_c)+\frac{T^2}{2}\alpha\gamma_c\right)\|x_1-x_2\|, \hspace{3mm}\forall \, x_{1},x_{2}\in \mathcal{D},
\end{align}
which shows the satisfaction of the Lipschitz continuity in discrete-time domain, using Cauchy-Schwartz inequalities for normed spaces. On the other hand, $\alpha$ itself by definition is the Lipschitz constant in continuous-time \cite{Marquez}.
\begin{equation}
\alpha = \sup_{x \in \mathcal{D}}\bigg\|\frac{\partial f}{\partial x}\bigg\| \triangleq \gamma_c.
\end{equation}
Simplifying the above yields to:
\begin{align}
\gamma_d = T\gamma_c+T^2\left(\overline{\sigma}(A)\gamma_c+\frac{\gamma_{c}^{2}}{2}\right).
\end{align}
\subsubsection{One-Sided Lipschitz Continuity}
Similar to the Lipschitz continuity, we can construct the one-sided Lipschitz continuity condition for the second order approximate model, using the properties of the inner-product spaces. After doing some linear algebra, we get:
\begin{align}
&\left<F_{T}^{a}(x_{1},u^{*})-F_{T}^{a}(x_{2},u^{*}), x_{1}-x_{2}\right> < \\ \notag
&\left[\left(T+\frac{T^2}{2}\overline{\sigma}(A)\right)\rho_c+\frac{T^2}{2}\overline{\sigma}(A)\gamma_c+\frac{T^2}{2}\overline{\sigma}(A)\gamma_c\rho_c\right]\|x_1-x_2\|^2,
 \hspace{3mm}\forall \, x_{1},x_{2}\in \mathcal{D},
\end{align}
which leads to the following expression for the discrete-time one-sided Lipschitz constant:
\begin{align}
\rho_d = T\rho_c+\frac{T^2}{2}\overline{\sigma}(A)(\rho_c+\gamma_c+\rho_c\gamma_c).
\end{align}
It is interesting to see that in this case, the discrete one-sided Lipschitz constant is not only a function of the continuous one-sided Lipschitz contact, but also a function of the two-sided continuous Lipschitz constant.
\subsection{Higher Order Approximations}
In most practical applications, first or second order discretization should be enough, specially since the sampling time can be selected small enough to ensure desired bounds on the approximation error. Furthermore, the expressions involving higher-order approximate models rapidly become very complicated. In particular, higher-order partial derivatives require tensor analysis of higher-orders.
In this section, we briefly discuss the third-order approximate model, and derive the analytical expression for the two-sided Lipschitz constant, which also serves as a hint to those of higher-order approximate models.
For the third order approximate model, under the ZOH assumption we get:
\begin{align}
x(k+1)&= \left(I+AT+\frac{T^2}{2}A^2+\frac{T^3}{6}A^3\right)x(k)\\ \notag
  &+Tf(x,u)+\frac{T^2}{2}\left(Af(x,u)+\frac{\partial f}{\partial x}Ax+\frac{\partial f}{\partial x}f(x,u)\right) \\ \notag
  &+\frac{T^3}{6}\Bigg(2A^2\frac{\partial f}{\partial x}x+2\bigg(\frac{\partial^2 f}{\partial x^2}A\bigg)x+\bigg(\frac{\partial^2 f}{\partial x^2}A\bigg)f(x,u)+2A\frac{\partial f}{\partial x}f(x,u)+2\bigg(\frac{\partial^2 f}{\partial x^2}f(x,u)\bigg)f(x,u) \\ \notag
  &+\bigg(\frac{\partial f}{\partial x}\bigg)^2 f(x,u)\Bigg).
\end{align}
Note that the second derivative of the vector field $f$ with respect to the state vector $x$ is not a Hessian matrix, but a tensor of order three.
For the two-sided Lipschitz continuity condition, after some tedious manipulations, the discrete Lipschitz constant is achieved as:
\begin{align}
  \gamma_d = T\gamma_c&+\frac{T^2}{2}\Biggl[\overline{\sigma}(A)\gamma_c+\bigg\|\frac{\partial f}{\partial x}\bigg\|\gamma_c+\bigg\|\frac{\partial f}{\partial x}\bigg\|\overline{\sigma}(A)\Biggr]\\ \notag
  &+\frac{T^3}{6}\Biggl[\bigg\|\frac{\partial^2 f}{\partial x^2}\bigg\|\overline{\sigma}(A)\gamma_c+2\bigg\|\frac{\partial f}{\partial x}\bigg\|\overline{\sigma}(A)\gamma_c \\ \notag
  &+2\bigg\|\frac{\partial f}{\partial x}\bigg\|^2\gamma_c+2\|f(x,u)\|\bigg\|\frac{\partial^2 f}{\partial x^2}\bigg\|\gamma_c
  +2\bigg\|\frac{\partial f}{\partial x}\bigg\|\overline{\sigma}^2(A)+2\bigg\|\frac{\partial^2 f}{\partial x^2}\bigg\|\overline{\sigma}(A)\Biggr].
\end{align}
Defining
\begin{align}
  \gamma_c &=\sup_{x \in \mathcal{D}}\bigg\|\frac{\partial f}{\partial x}\bigg\|, \\ \notag
  \beta &=  \sup_{x \in \mathcal{D}}\bigg\|\frac{\partial^2 f}{\partial x^2}\bigg\|, \\ \notag
  M &= \sup_{x \in \mathcal{D}} \|f(x,u)\|,
\end{align}
the above is simplifies to
\begin{align}
\gamma_d = T\gamma_c&+T^2\left(\overline{\sigma}(A)\gamma_c+\frac{\gamma_{c}^{2}}{2}\right) \\ \notag
&+\frac{T^3}{6}\Bigg[2\beta\overline{\sigma}(A)+\bigg(\beta\overline{\sigma}(A)+2\beta M+2\overline{\sigma}^2(A)\bigg)\gamma_c
+2\overline{\sigma}(A)\gamma_{c}^{2}+2\gamma_c^3\Bigg].
\end{align}
\section{Conclusions}

The preservations of the (one-sided) Lipschitz continuity condition was studied for nonlinear systems undergoing a ZOH based approximate discretization. It was shown that the condition can still be established for approximate discrete model, with Lipschitz constants as a function of the continuous Lipschitz constant, and sampling time and induced norms of the system matrices and its Jacobian.


\bibliographystyle{IEEEtran}
\bibliography{References}


\end{document}